\documentstyle[12pt]{article}
\makeatletter
\@addtoreset{equation}{section}
\makeatother

\def\ee{\end{equation}}

\title{Nt-yet}
\begin{document}

\begin{flushright}
\vspace{1mm}
FIAN/TD/
 june {2002}\\
\end{flushright}

\vspace{1cm}

\begin{center}
{\large\bf DOES CLASSICAL MECHANICS ALWAYS ADEQUATELY DESCRIBE "CLASSICAL PHYSICAL REALITY" ?}
\vglue 0.6  true cm
\vskip1cm
{\bf V.E.~Shemi-zadeh\footnotemark[1]} \footnotetext[1]{Internet adress: www.shemizadeh.narod.ru}
\vglue 0.3  true cm

I.E.Tamm Department of Theoretical Physics, Lebedev Physical Institute,\\
Leninsky prospect 53, 119991, Moscow, Russia
\vskip2cm
\end{center}

\begin{abstract}
The article is dedicated to discussion of irreversibility and foundation of statistical mechanics "from the first principles". Taking into account infinitesimal and, as it seems, neglectful 
for classical mechanics fluctuations of the physical vacuum, makes a deterministic motion of unstable 
dynamic systems is broken ("spontaneous determinism breaking", "spontaneous 
stochastization"). Vacuum fluctuations play part of the trigger, starting the powerful mechanism 
of exponent instability. The motion of the dynamic systems becomes irreversible and stochastic. 
Classical mechanics turns out to be applicable only for a small class of stable dynamic systems 
with zero Kolmogorov-Sinay entropy $h=0$. For alternative "Stochastic mechanics" there are 
corresponding equations of motion and Master Equation, describing irreversible evolution of the 
initial distribution function to equilibrium state. 
\end{abstract}
$$ $$

In 1893 A.Poincare published a short article "Le mechanisme et l'experience" [1]. The article 
sum up a violent discussion on the problem of time irreversibility between L.Boltzmann on the 
one part and J.Loschmidt and E.Zermelo on the other part. Poincare's conclusion was, naturally, 
categorical and far-seeing. He concluded that paradoxes of irreversibility mean, evidently, that 
there is a serious failure of mechanics, and solving it will take revision of some basic rules, 
building up some theory, more general than the mechanics. The community of physicists, 
evidently, has not accepted these true in their nature conclusions. Anyway W. Tomson (Lord 
Kelvin) in the end of the 19 century in his famous report summing up the development of 
theoretical physics does not mention these important conclusions of Poincare, though speaking 
about notorious two "clouds" in the sky of physics: experiment of Michelson-Morley and 
difficulties of description of radiation of absolutely black body. 

The problem of reversibility-irreversibility has not by now found adequate solution. 
V.L.Ginzburg in his famous program article "What Problems of Physics and Astrophysics Seems 
Now to be Especially Important and Interesting" [2] sets it among three most important "great 
problems".

General notion about the state of the matter, about different points of view of the problem, 
possible ways of solving it can be found in the following works of [3].

Nowadays it is principally clear that appearance of irreversibility is somehow connected 
with instability of classical dynamical systems. This point of view is supported by most of the 
researchers.

Below we will consider the central question: how this mechanism of local instability is 
started, how the deterministic motion is broken, how stochasticity appears in the behavior of 
dynamic systems, how the behavior of dynamic systems becomes time irreversible. Meanwhile 
the area of correctness of classical mechanics becomes more narrow, as well as the areas of 
adequate description of classical dynamic systems. We will come to the need of generalization of 
classical mechanics and formulate the basic equations of the new stochastic mechanic. In other 
words we will offer the solution of the paradox of reversibility- irreversibility. 

As introduction let us consider the behavior of knowingly classical material point with mass 
of m=1 g in three situations: free state, in potential pit and on the top of potential hill. 

a)      Free mass is in the free state. What is its further behavior at strict consideration? 
By strict consideration we mean consideration of the mass's behavior not in ideal 
mathematic vacuum, but in real physical one under the influence of vacuum fluctuations 
to our classical mass. 

We suppose that, as in quantum mechanics, fluctuations of coordinate  $\delta x$ and 
conjugated momentum $\delta p$ (to make it simple we consider only one-dimensional 
case), for which the dispersions meet the relation
$$<\delta x^2><\delta p^2>=\hbar^2/4 .\eqno(1)$$

Due to isotropy the mean values $<\delta x>=0$,$<\delta p>=0$.

For instant value of coordinate for a small period of time $\tau $ we can write:
$$ \delta x_\tau = \frac{\delta p_\tau}{m} \tau .\eqno(2)$$

Taking the square of expression (2), taking mean value and excluding $<\delta p_\tau ^2>$ by 
means of (1) we get
$$<\delta x_\tau^2>=\frac{\hbar}{2m}\tau .\eqno(3) $$

It coincides with the formula for mean square root deviation of  Brownian 
particle in the medium with diffusion coefficient $\hbar/4m$:
$$<\delta x^2>=2D\tau .$$

In the terms of stochastic calculs [4] we can record the equations of motion of 
the classical mass:
$$dx=\left( \frac{\hbar}{2m}\right) ^{1/2}dw, \eqno(4)$$
where {\it dw} is the differential of Wiener process with attributes $<dw>=0$, \ \ $<dw^2>=dt$.

So our mass, despite the stipulations of the classical mechanics does not keep the 
quiescence, but is moving chaotically, diffunding according to (4).
But is there any ground for refusal from the principles of classical mechanics?
The mean square root deviation of mass m=1g from the initial position for the age of the 
Universe (13 billion years) will make
$$ <\delta x^2>^{1/2} \sim 10^{-5} \ \ cm, $$ 

That is infinitesimal value, not giving ground for revision of classical mechanics.

b)      The next example is the same mass m in potential pit of harmonic 
oscillator. 
$$V(\delta x)=\frac{1}{2}m\omega ^2 \delta x^2. $$ 

The coordinate and momentum as we suppose fluctuate here too. The mean value of 
energy
$$<E>=\frac{1}{2m}<\delta p^2>+\frac{1}{2}m \omega ^2 <\delta x^2>, $$

Considering (1) has minimum value at
$$<\delta x^2>=\hbar/2m\omega, \ \ \ \ <\delta p^2> =\frac{1}{2}\hbar m \omega , $$

That finally giving the formula we know
$$ <E>=\frac{1}{2}\hbar \omega .$$

And here the mean energy and oscillations amplitude are infinitesimal and are of no 
practical interest.

c)      The last example is the mass in the state of unstable equilibrium on the top of 
potential 
$$V(\delta x)=-\frac{1}{2} m\lambda^2 \delta x^2 . $$

For $\delta x_\tau$ we have from the motion equations
$$ \delta x_\tau =\delta x_0 ch (\lambda \tau) + \frac{\delta p_0}{\lambda m} sh({\lambda \tau}) . $$

Here for times $\tau >> \lambda ^{-1} $:
$$ <\delta x_\tau^2>=\frac{\hbar}{4m\lambda}e^{2\lambda \tau} . \eqno(5)$$

Here as we see the situation is principally different. Infinitesimal noise plays the part of 
trigger and start the powerful mechanism of instability. The material point spontaneously falls 
from its unstable position of equilibrium to the left or to the right.
 
So we are ready to consideration of the main matter, the spontaneous breaking of 
deterministic motion of unstable dynamical systems, stochastization and occurrence of 
irreversibility.

Let us consider some dynamic system with Hamiltonian ${\cal H}(x,p)$.

Motion equations
$$\frac{dx_i}{dt}=\frac{\partial {\cal H}}{\partial p_i},\ \  \frac{dp_i}{dt}=-\frac{\partial {\cal H}}{\partial x_i} . \eqno(6)$$ 

Liuville equation
$$\frac{\partial \rho }{\partial t}+{\cal L}\rho =0, $$

Where ${\cal L}$ is the Liuville operator.

Stability of some trajectory can be estimated by the motion of trajectory close to the initial 
one ${\overline x_i}=x_i+\delta x_i , {\overline p_i}=p_i+\delta p_i$.
Or by the solutions of linearized equation for variations  $\delta x_i $ and $\delta p_i$:
$$\delta {\dot x_i} = \sum_j \frac{\partial ^2 {\cal H}}{\partial p_i \partial p_j} \delta p_j, \ \ 
\ \delta {\dot p_i}=-\sum_j \frac{\partial ^2 {\cal H}}{\partial x_i \partial x_j} \delta x_j ,$$ 

In the matrix form
$$
\left(
\begin{array}{c}
 \delta {\dot p}\\
 \delta {\dot x}
\end{array} \right) =
\left(
\begin{array}{cc} 0 & A\\B&0 \end{array}\right)\left( \begin{array}{c} \delta p \\ \delta x \end{array} \right),
A_{ij}=-\frac{\partial ^2{\cal H}}{\partial x_i \partial x_j} , B_{ij}=\frac{\partial ^2 {\cal H}}{\partial p_i \partial p_j}.
$$

The matrix to the right can be led by means corresponding transformation $2 \times 2$ to the 
view with matrixes on the diagonal. The matrixes can be of three types specified 
below (a,b,c).
$$\left(
\begin{array}{cc}
0 & 0\\
1/m_i & 0
\end{array}
\right)
,
\left(
\begin{array}{cc}
 0 & -m_i \omega_i^2\\
 1/m_i& 0
\end{array} \right)
, \left(
\begin{array}{cc}
 0 & m_i \lambda_i^2\\
 1/m_i & 0
\end{array}
\right) ,
$$

Actually, the first matrix shows that the corresponding degree of freedom does not interplay 
with the other ones, the corresponding index of Lepunov is equal to zero, the degree of freedom 
is stable (case a).

The second matrix (case b) also corresponds to the stability of the freedom degree, which 
oscillates with infinitesimal amplitude near the initial trajectory.

The third matrix corresponds to the case of exponential instability. Here at motion of this 
freedom degree every moment there occurs a spontaneous jump into infinitesimal trajectory, 
there is Brownian motion of the showing points on the neighbor trajectories, the mechanism of 
exponential deviation of trajectories switches on, that leads to breaking of deterministic motion 
and to stochastization.

Similar process takes place on all the instable degrees of freedom. In this case the equation 
for variations (letting alone the index)
$$
\left(
\begin{array}{c}
\delta {\dot p}\\
\delta {\dot x}
\end{array} \right)
=\left(
\begin{array}{cc}
0 & m\lambda ^2\\
1/m & 0 
\end{array}
\right)
\left(
\begin{array}{c}
\delta p\\
\delta x
\end{array}
\right),
$$
Or
$$
 \delta {\dot p}=m\lambda ^2 \delta x , \ \ \delta {\dot x}=\frac{1}{m}\delta p.
$$

Performing analogous to (1)-(4) calculations we get:
$$<\delta x_i^2> = \frac{\hbar}{2m}\delta \tau ,\ \  <\delta p^2>=\frac{1}{2}\hbar m \lambda ^2 \delta \tau , $$

Where
$$ \frac{\hbar}{4m},\ \  \frac{1}{4}\hbar m \lambda ^2 \  -$$

are the diffusion coefficients according to coordinate and impulse.
Inputting as before the stochastic differentials ${\overline d}x$ and ${\overline d}p$, we have 
$${\overline d}x=\left( \frac{\hbar}{2m} \right) ^{1/2}dw_x , \ \ \ \ {\overline d}p=\left( \frac{\hbar m \lambda ^2}{2} \right) ^ {1/2}dw_p \ ,$$

We have to revise Hamilton motion equations (6) as stochastic Hamilton equations:
$$ dx_i=\frac{\partial {\cal H}}{\partial p_i} dt+(\hbar/2m)^{1/2}dw_x , \ \ \ dp_i=-\frac{\partial {\cal H}}{\partial x_i}dt + (\hbar m \lambda ^2 /2 )^{1/2} dw_p .\eqno(7)$$

These are typical Itoh's stochastic differential equations. They describe stochastic and 
irreversible motion. Liuville's equation appears to becomes irreversible:
$$\frac{\partial \rho }{\partial t}+{\cal L}\rho = \frac{\hbar}{4}D\rho , \eqno(8)$$

Where ${\cal L}$ - is Liuville's operator and  D- the diffusion operator. 
This equation of  Fokker-Plank type describes irreversible evolution of the initial 
distribution function $\rho (0)$ to the equilibrium state.
The diffusion operator  in transformed (normal) coordinates is:
$$ D=\sum_i \left( \frac{\partial ^2 {\cal H}}{\partial p_i^2} \frac{\partial ^2}{\partial x_i^2} + \frac{\partial  ^2 {\cal H}}{\partial x_i^2} \frac{\partial ^2}{\partial p_i^2} \right), $$
Where the summing is on all the unstable degrees of freedom.

Finally, returning to the initial coordinates by reverse transformation we have instead of 
reversible Liuville equation an irreversible equation of the Fokker-Plank type (8) with non-negative
 defined diffusion operator D-containing second derivatives on coordinates and 
momentums. 

So we have modified classical mechanics and  came to the new, more general theory, i.e.
"Scholastic mechanics". All the dynamical systems with the unstable  trajectories (where the 
 entropy   $h>0$) cannot be adequately described by the 
classical mechanics. These systems are irreversible and have in their motion is stochastic random 
elements and are described by the above-mentioned stochastic mechanics. 

Equation (8) can be named Master Equation. It, at stability of all the degrees of freedom 
($h=0$) transforms into Luiville equation, in the same way with the stochastic 
equations of Hamilton (7) transforms into ordinary Hamilton equations of motion (6). These 
are the general features of solution of the problem of reversibility-irreversibility and the answer 
to the question about where stochastic and irreversible motion of dynamical systems is taken 
from. The answer to the question set in the title of the article must be as follows: "No, not 
always, but very seldom, classical mechanics adequately describes only a narrow class of stable 
dynamic systems".
$$ $$
The present work is performed with all-round support of the Center for Advanced Research: 
Theoretical and mathematical physics. The author thanks Smurniy Ye.D. for assistance to 
preparation of the article to publication. Also the author would like to thank Kutuzova T.S. for 
fruitful discussions.


\begin{thebibliography}{10}
\bibitem{1} Poincare H. - Rev. Methaphys. et Morale, 1893, vol. 1, p. 534-537.

\bibitem{2} Ginzburg V.L. Uspehi Fizicheskih Nauk, 1999, 169, ü 4,  p.420-441.

\bibitem{3} Prigogine I. - From Beginning to Becoming: time and complexity in the physical 
sciences, W.H.Freeman and Company, San Francisco, 1980, 217.  
\bibitem{6}  Gardiner C.W. - Handbook of Stochastic Methods for Physics, Chemistry and the 
Natural Science, Springer, Berlin, 1985, 526 p.

\end{thebibliography}
\end{document}